\documentclass[aps,prb,reprint,superscriptaddress]{revtex4-1}
\usepackage{graphicx}
\usepackage{array}
\usepackage{units}
\usepackage{hyperref}

\def\NGOS {NdGaO$_{3}$/SrTiO$_{3}$}
\def\LAOS {LaAlO$_{3}$/SrTiO$_{3}$}
\def\LAO {LaAlO$_{3}$}
\def\STO {SrTiO$_{3}$}
\def\NGO {NdGaO$_{3}$}

\begin{document}

%\preprint{}

\title{Observation of strontium segregation in LaAlO$_{3}$/SrTiO$_{3}$ and NdGaO$_{3}$/SrTiO$_{3}$ oxide heterostructures by X-ray photoemission spectroscopy}


\author{Uwe Treske}
%\thanks{}
\author{Nadine Heming}
\author{Martin Knupfer}
\author{Bernd B\"uchner}
\affiliation{Institute for Solid State Research, IFW-Dresden, P.O. Box 270116, DE-01171 Dresden, Germany}

\author{Stefan Krause}
\affiliation{Helmholtz-Zentrum Berlin, BESSY, Albert-Einstein-Str. 15, 12489 Berlin, Germany}

\author{Emiliano Di Gennaro}
\author{Umberto Scotti di Uccio}
\author{Fabio Miletto Granozio}
\affiliation{CNR-SPIN and Dipartimento di Fisica, Complesso Universitario di Monte S.Angelo, Via Cintia, 80126 Naples, Italy}

\author{Andreas Koitzsch}
\email[]{a.koitzsch@ifw-dresden.de}
\affiliation{Institute for Solid State Research, IFW-Dresden, P.O. Box 270116, DE-01171 Dresden, Germany}

\date{\today}

\begin{abstract}
{LaAlO$_{3}$} and {NdGaO$_{3}$} thin films of different thickness have been grown by pulsed laser deposition on TiO$_2$-terminated {SrTiO$_{3}$} single crystals and investigated by soft X-ray photoemission spectroscopy. The surface sensitivity of the measurements has been tuned by varying photon energy $h\nu$ and emission angle $\Theta$. In contrast to the core levels of the other elements, the Sr $3d$ line shows an unexpected splitting for higher surface sensitivity, signaling the presence of a second  strontium component. From our quantitative analysis we conclude that during the growth process Sr atoms diffuse away from the substrate and segregate at the surface of the heterostructure, possibly forming strontium oxide.
\end{abstract}

\maketitle

\section{Introduction}
In recent years pulsed laser deposition (PLD) has proved to be a powerful tool for stoichiometric epitaxial growth of a target material on top of a single crystal substrate. In combination with reflection high energy electron diffraction (RHEED) an atomically controlled layer-by-layer deposition is possible. This enables the manufacturing of epitaxial heterostructures exhibiting intriguing physical and electronic properties.\cite{Ohtomo2002, Brinkman2007, Aruta2010}

At the interface between two insulating metal oxides, \LAO\ and \STO\ for example, a thickness threshold for an insulator to metal phase transition has attracted much interest. A highly mobile two dimensional electron gas (2DEG) is formed at the interface when 4 or more unit cells (uc) of epitaxial \LAO\ are deposited on a single crystal, TiO$_2$-terminated \STO\ substrate.\cite{Ohtomo2004,Thiel2006} The origin of this unexpected behavior has been so far the subject of passionate debates.
The "polar-catastrophe" scenario is believed by many scientists to catch most of the physics of this system. Within such model the electrostatic potential rises steadily with the growth of an increasing number of polar \LAO\ layers, until the heterostructure accesses a new electrostatic ground state where a so-called “electronic reconstruction” process, transferring electronic charges from \LAO\ to \STO, takes place.\cite{Ohtomo2004, Nakagawa2006}
However, it has been argued that oxygen vacancies may affect the interface conductivity,\cite{Kalabukhov2007,Siemons2007, Chen2011} which in fact  depends on the oxygen partial pressure during growth and post-deposition annealing procedures. Finally, deviations from an abrupt interface due to cation intermixing at the \LAOS\ interface has also been proposed as a possible source of chemical doping, giving rise to the observed interface conductivity. Such intermixing effects were found, e.g. by X-ray diffraction (XRD) \cite{Willmott2007}, medium energy ion scattering \cite{Kalabukhov2009} and photoemission spectroscopy \cite{Qiao2010} for both the A-site (La $\leftrightarrow$ Sr) and B-site sublattice (Al $\leftrightarrow$ Ti). \cite{Chambers2010,Chambers2011,Vonk2007,Vonk2012} Recent scanning transmission electron microscopy and electron energy loss spectroscopy (STEM/EELS) data performed on samples similar to the ones employed in this work have posed nevertheless stringent upper limits to the amount of La cations crossing the interface. \cite{Cantoni2012}

X-ray photoemission spectroscopy (XPS) is a surface sensitive technique that enables in principle the investigation of all three scenarios mentioned above. Among the challenges of the method are the distinction of interface and bulk signal and the strong damping of the interface signal by the overlayer. We investigated the cation core level spectra of \LAOS\ and \NGOS\ samples by tuning the surface sensitivity. \NGOS\ shares with \LAOS\ a perovskite structure, an insulating nature of the single building blocks, a polar/non-polar character, and a critical thickness of four unit cells for the onset of conductivity. Furthermore, it also possesses transport properties that are similar to \LAOS. \cite{DiGennaro2013} The electrical properties of the \NGO/\STO\ system have been investigated recently as a function of the growth conditions.\cite{Gunkel2013}

Here we show that an unexpected second strontium component, that we attribute to surface segregation of Sr cations, clearly emerges from collected spectra. Such phenomenon is frequently found in perovskites such as La$_x$Sr$_{1-x}$MnO$_3$, \cite{Dulli2000} \STO, \cite{Clausthal-Zellerfeld2002} SrTi$_{1-x}$Fe$_x$O$_{3-\delta}$ \cite{Jung2012} and in the TiO$_2/$SrTiO$_3$ system. \cite{Radovic2011,Ciancio2012} Also the role of Sr-vacancies at the \LAO/\STO\ interface is under discussion. \cite{Gunkel2012} The Sr segregation effect occurs for all heterostructures and film thicknesses addressed in our investigation.

%----------------------------------------------------------------------------------------------------------------------
\section{Experimental}
Room temperature soft X-ray photoemission spectroscopy measurements have been carried out at the UE52-PGM Beamline of the Berlin Synchrotron Facility (BESSY). Photon energies ($h\nu$) varying from  200\,eV to 1200\,eV were chosen to control the surface sensitivity of the measurements. Spectra have been recorded with a Scienta SES 4000 energy analyzer and a total energy resolution smaller than $\Delta E = 250$\,meV, depending on $h\nu$. For quantitative interpretations the photon energy dependent inelastic mean free paths (IMFP) $\lambda$ of the photoelectrons have been calculated by the semi-empirical TPP-2M model\cite{TPP-2M} that considers the material specific density and band gap of the grown layer. The corresponding cross sections and asymmetry parameters for the core levels are given by Yeh and Lindau. \cite{sigma}

\LAOS\ and \NGOS\ samples were grown by RHEED-assisted PLD technique on nominally single terminated \STO\ substrates that were purchased from TSST BV. The deposition conditions were as follows: substrate temperature $T = 730\,^\circ$C, oxygen partial pressure $P_{\mathrm{O_2}} = 10^{-2}$\,mbar, and laser fluence $F = 1.5\,\mathrm{J} / \mathrm{cm}^2$ on the target. The target - substrate distance was 40 mm. A relatively high oxygen partial pressure was adopted in order to decrease the risk of incorporating oxygen vacancies into \STO\ during the growth process. This method has proved to yield high quality metallic samples, above threshold thickness, when proper growth conditions are adopted. \cite{DiGennaro2013,Aruta2010,Aruta2012} A slow cooldown to room temperature was performed after growth by keeping the oxygen pressure unchanged.

Samples were transported in air and no further cleaning steps were applied prior to experiment.

%------------------------------------------------------------------------------------------------------------
%----------------------------Results--------------------------------------------------------------------------------

\section{Results and Discussion}
\begin{figure}
\includegraphics[width=0.99\columnwidth]{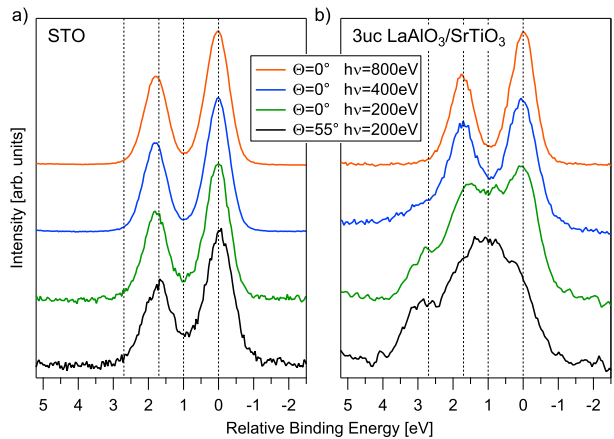}%
\caption{Background subtracted and normalized Sr~$3d$ photoemission core level spectra of a \STO\ substrate (a) and the 3\,uc \LAOS\ sample (b) measured at different photon energies $h\nu$ and emission angles $\Theta$. Surface sensitivity is enhanced for lower photon energies and higher emission angles. The energy scale is referred to the Sr~$3d_{5/2}$ maximum in order to neglect charging effects.\label{Fig1}}
\end{figure}

Fig.\,\ref{Fig1} shows the background corrected and normalized Sr~$3d$ core levels for pure \STO\ (a) and 3\,uc \LAOS\ (b) measured with varying photon energy and emission angle. The Sr~$3d$ line consists of a doublet due to spin-orbit splitting (Sr~$3d_{3/2}$, Sr~$3d_{5/2}$). Charging-related energy shifts have been encountered during the measurements. We will refer therefore in the following to the binding energy (BE) shifts with respect to the Sr~$3d_{5/2}$ peak maximum, rather than to absolute BE values. The surface sensitivity of the experiment can be tuned by changing the photon energy and the emission angle. Lower photon energies and higher emission angles increase the surface contribution to the overall signal and decrease the effective IMFP $\lambda_{\mathrm{eff}}$.

The \STO\ spectra show two single peaks with little dependence of the shape on measuring parameters, although in the most surface sensitive conditions ($\Theta = 55^{\circ}$, $h\nu = 200$\,eV) a slight broadening might be present. The spectra collected for \LAOS, on the contrary, feature a strong dependence on $\lambda_{\mathrm{eff}}$. Whereas the $h\nu = 800$\,eV spectrum resembles the one of \STO, at lower photon energies and higher emission angle an increasing high BE component is found, that dominates the spectrum for $\Theta = 55^\circ$, $h\nu = 200$\,eV.

Such behavior is observed on the Sr~$3d$ doublet of all the \LAOS\ and \NGOS\ samples, irrespective of polar film thickness. The core levels of the other cations do not show, instead, any comparable dependence on the measuring conditions. 

The change of the Sr~$3d$ line in Fig.\,\ref{Fig1} is most naturally associated with the appearance of a second Sr component, which is chemically inequivalent with respect to the Sr$^{2+}$\,cations populating the perovskite A-site in \STO. Alternative possibilities, as a strong band bending in \STO\ near the interface, would also cause a $\lambda_{\mathrm{eff}}$ dependence of
the spectra. Such hypothesis is nevertheless inconsistent with the absence of a comparable broadening for the other core levels. Furthermore, band bending would lead to a continuous shift of the peak positions of the Sr spectra as a function of $\lambda_{\mathrm{eff}}$ between the two extreme BE values, for which no evidence was found in our data.

%------------------------------------------------------------------------------------------------------------------

In order to extract quantitative information from the spectra we implemented a global fitting scheme. All the spectra for a given sample have been fitted with two doublets and the following constraints: The intensity ratio for the spin-orbit components has been set to 3 : 2, equal FWHM (full width half maximum) for doublet lines were imposed and the same energy separation between the two doublets was globally claimed for all measurements of a sample. 

\begin{figure}
 \includegraphics[width=0.99\columnwidth]{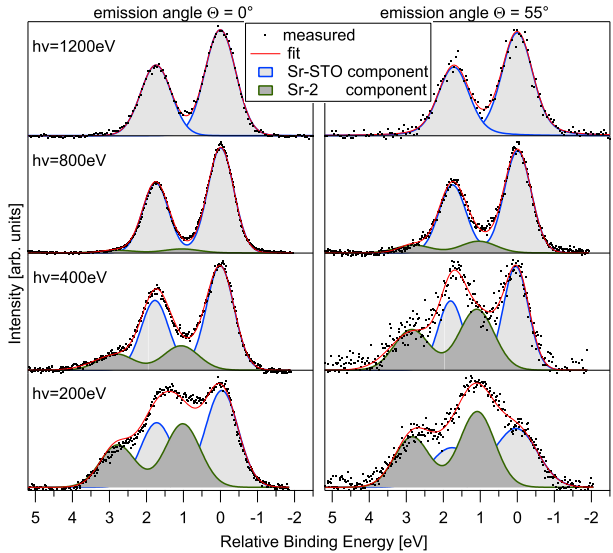}%
 \caption{Fitted Sr~$3d$ core level photoemission spectra of a 3\,uc \LAOS\ sample, that was measured at different photon energies $h\nu$ and emission angles $\Theta$. For higher emission angles or lower photon energies (higher surface sensitivity) the second Sr component is enhanced.\label{Fig2}}
\end{figure}

The results are shown in Figure \ref{Fig2} and Table \ref{tab:fit}. The fit is of good quality. For all three samples the same parameters satisfy (and therefore validate) the described procedure. The energy separation between the substrate and second Sr component is about 1\,eV. 
The larger FWHM of the photoemission profile from the high BE component, called Sr-2 in the following, suggests a higher degree of disorder of Sr-2 cations with respect to the Sr-STO cations residing in crystalline \STO. 

\begin{table*}
\centering
\caption{Global fit parameters for the three different heterostructures.} \label{tab:fit}
\begin{tabular}{|l|c|c|c|}
\hline
Parameter												& 3uc \LAO					& 3uc \NGO					& 6uc \NGO \\
\hline
Energy separation	(eV)						& 1.06	& 0.95	& 1.00	\\
FWHM(Sr-STO)	(eV)								& 0.9		& 0.9		& 0.9		\\
FWHM(Sr-2)	(eV)									& 1.2		& 1.3		& 1.3	\\
\hline
\end{tabular}
\label{tab:Sr_Fit_Para}
\end{table*}

It is obvious in Figure \ref{Fig2} that the integrated intensity of the Sr-2 component increases for lower $h\nu$ and larger $\Theta$ compared to the integrated intensity of Sr-STO, proving that Sr-2 cations are located above the \STO\ substrate.

Still, a number of more challenging issues about the \mbox{Sr-2} component remain open: are the Sr-2 cations lying in-between \LAO\  and \STO, are they homogeneously intermixed in \LAO\ or do they segregate above \LAO? Which is their amount? Which is their source and which is the origin of the chemical shift?

In Fig.\,3(a)  the integrated intensity ratios (IIRs) between the Sr~$3d$ and La~$4d$ profiles in \LAOS\ are plotted vs $\lambda_{\mathrm{eff}}$ for both Sr components. Each data point corresponds to a given photon energy and emission angle that is converted into $\lambda_{\mathrm{eff}}$. The data were corrected for core level and photon energy dependent cross sections and asymmetry parameters.\cite{sigma} As expected, the Sr-STO IIR decreases exponentially for lower $\lambda_{\mathrm{eff}}$, due to the rising photoelectron damping by the \LAO\ overlayer and the simultaneously increasing contribution of La. In contrast to this the Sr-2 IIR increases for lower $\lambda_{\mathrm{eff}}$. Once again, a similar behavior is observed for the \NGOS\ samples shown in Fig.\,3(b) and 3(c): the Sr-STO IIR (measured with respect to Ga~$3p$) increases with increasing $\lambda_{\mathrm{eff}}$, while Sr-2 clearly shows an opposite trend.
If the Sr-2 cations were lying in between \LAO\ (\NGO) and \STO, their IIR would show the same trend as Sr-STO. A homogeneous intermixing of Sr-2 in \LAO\ (\NGO) would cause a constant IIR vs $\lambda_{\mathrm{eff}}$. Only a top surface position of the Sr-2 cations entails the observed increase of IIR with decreasing $\lambda_{\mathrm{eff}}$.

The measured IIR profiles can be compared to a model that assumes an abrupt interface and an additional overlayer containing the Sr-2 species (see Fig.\,3 inset). Within this scenario, the IIR values for Sr-STO and Sr-2 are expressed as a function of the thicknesses by Eq.\ (1) and (2):
\begin{equation}
\frac{I_{\mathrm{Sr-STO}}}{I_{\mathrm{La}}} = \frac{I^0_{\mathrm{Sr-STO}} \times e^{\nicefrac{-D_{\mathrm{LAO}}}{\lambda_{\mathrm{eff}}}}}{I^0_{\mathrm{La}}  \times (1-e^{\nicefrac{-D_{\mathrm{LAO}}}{\lambda_{\mathrm{eff}}}})},
\label{eq:1}
\end{equation}
\begin{equation}
\frac{I_{\mathrm{Sr-2}}}{I_{\mathrm{La}}} = \frac{I^0_{\mathrm{Sr-2}} \times (1-e^{\nicefrac{-D_{\mathrm{Sr-2}}}{\lambda_{\mathrm{eff}}}})}{I^0_{\mathrm{La}}  \times (1-e^{\nicefrac{-D_{\mathrm{LAO}}}{\lambda_{\mathrm{eff}}}}) \times e^{\nicefrac{-D_{\mathrm{Sr-2}}}{\lambda_{\mathrm{eff}}}}}.
\label{eq:2}
\end{equation}

\begin{figure}
  \includegraphics[width=0.8\columnwidth]{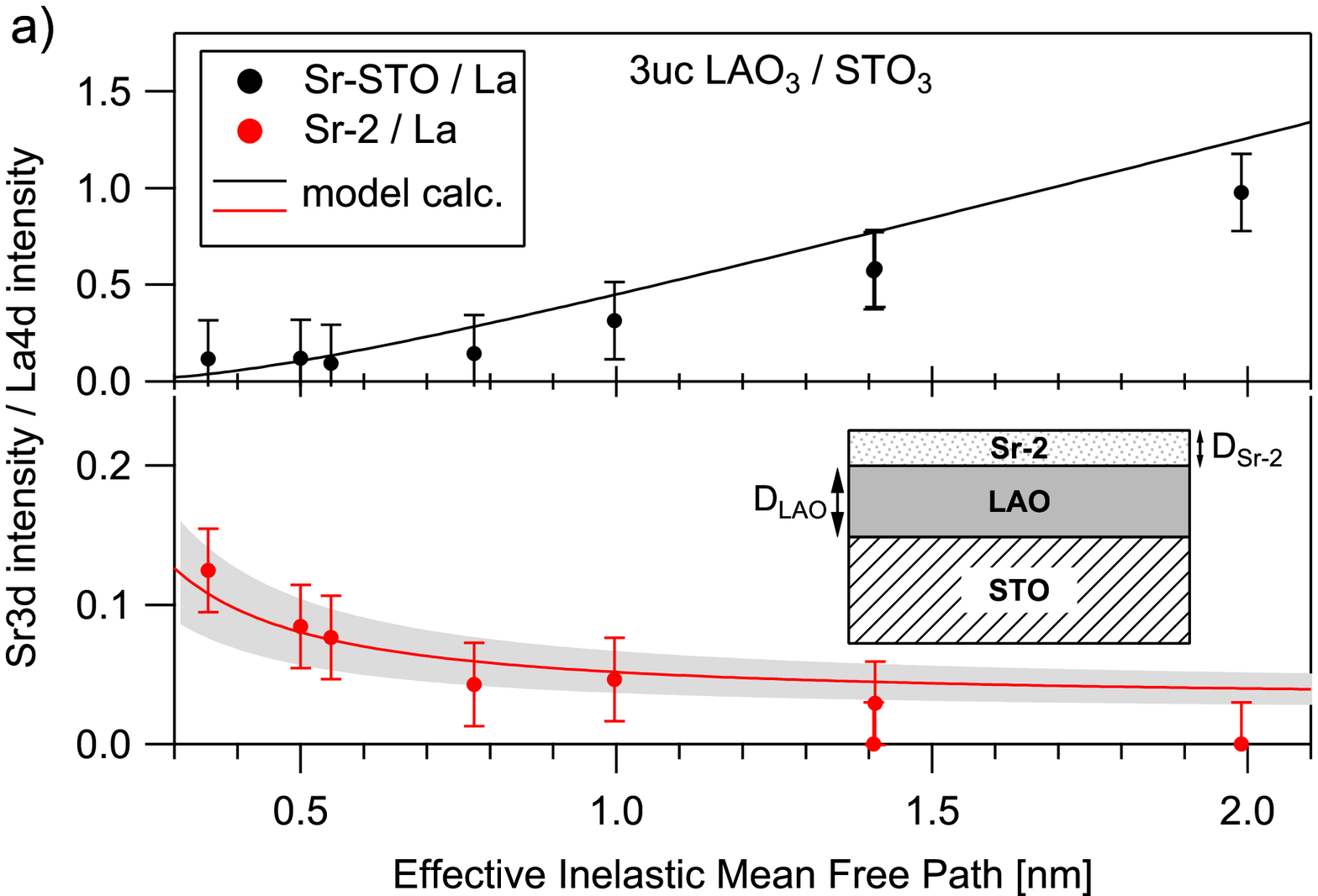}%

 \includegraphics[width=0.8\columnwidth]{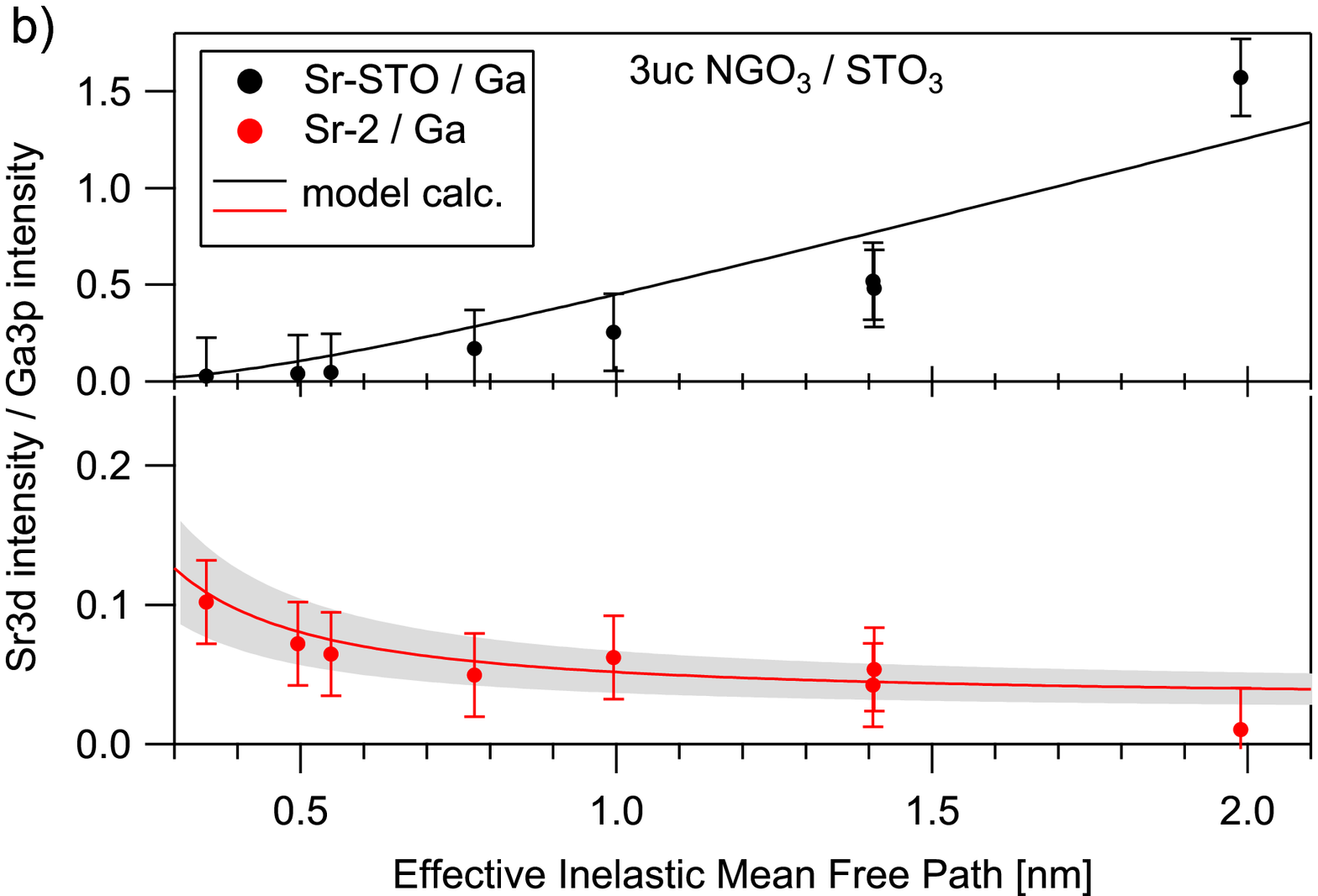}%

 \includegraphics[width=0.8\columnwidth]{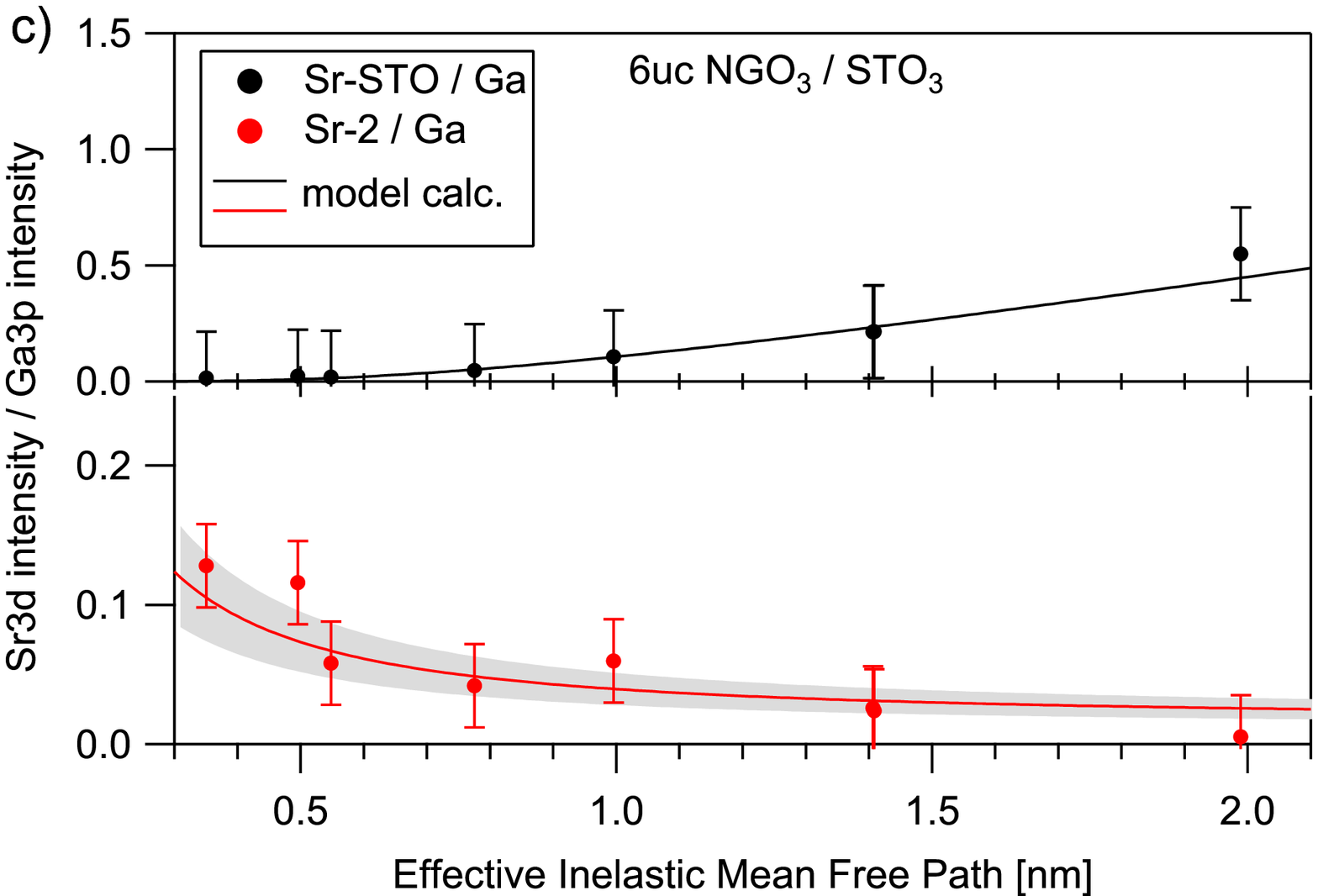}%
 \caption{Sr\,$3d$/La\,$4d$ (Ga\,$3p$) intensity ratio of 3\,uc \LAOS\ (a), 3\,uc \NGOS\ (b), and  6\,uc \NGOS\ (c) vs.\ effective inelastic mean free path of the photoelectrons. Lines and gray area: Model calculations with  $D_{\mathrm{Sr-2}}=(0.035 \pm 0.01)$\,nm. See main text for details. Inset: Sketch of the heterostructure including the inferred position of the Sr-2 component.\label{Fig3}} 
\end{figure}

Here $I$ are the integrated intensities, directly extracted by fitting the data for Sr-STO and Sr-2 and La.\cite{Note} $I^0$ reflects the concentrations and cross sections of the different elements. In order to reduce the number of free parameters we set $I^0_{\mathrm{Sr-STO}}$ = $I^0_{\mathrm{Sr-2}}$, i.e. we assume the unknown atomic concentrations of Sr-2 to be equal to the Sr-STO case.  D is the thickness of the layers (see Fig.\,3 inset). 
The solid lines in Figure \ref{Fig3} represent the calculated profiles of Eq.\ (\ref{eq:1}) and (\ref{eq:2}) for a 3\,uc \LAO\ and 3\,uc and 6\,uc \NGO\ layer ($D_{\mathrm{3\,uc}} = 1.2$\,nm, $D_{\mathrm{6\,uc}} = 2.4$\,nm). 
The  thickness of the Sr surface layer $D_{\mathrm{Sr-2}}$ describing the datasets best is in the range of  $D_{\mathrm{Sr-2}} = 0.03 - 0.04$\,nm.  

The model describes the data successfully and confirms that the Sr-2 cations lie predominantly at the surface of the heterostructures. $D_{\mathrm{Sr-2}}$ appears to be much smaller than a \STO\ unit cell and all $\lambda_{\mathrm{eff}}$. For this situation the model employed in (1) and (2), which rests on the IMFP formalism, should be considered as an attempt to extract the order of magnitude of the top Sr concentration rather than exact quantization. The latter comes out as only a fraction of a single \STO\ unit cell, which hints to partial coverage or island formation. The model itself assumes a homogeneous Sr-2 layer, realized if the Sr is incorporated in the terminating \LAO\ layer or by finely dispersed Sr-based molecules or clusters. However, to discriminate between these situations or other forms of island growth is not possible based on the data in Fig. 3 with certainty and remains, in essence, a task for future studies. 
Interestingly $D_{\mathrm{Sr-2}}$, i.e. the amount of Sr-2, appears to be independent of the chemical nature of the polar overlayer (\LAO\ or \NGO) and of its thickness.
Its observation requires low photon energies, rarely used for core level studies. This is probably the reason why this interesting phenomenon escaped the attention of previous investigations,\cite{Chambers2010,Takizawa2011} although some broadening of the Sr~$3d$ has been reported occasionally.\cite{Drera2013} 

We now turn to the two final and most intriguing open questions: which is the origin of the extra Sr-2 component and what determines the shift in BE? The extra Sr might well migrate from the bulk of \STO, that can be considered for our purposes as an “infinite Sr reservoir”. In this context, the driving force for Sr migration could either be an intrinsic non-stoichiometry of the single crystal or possibly the energy gain of a surface redox reaction of Sr in oxidizing conditions. As an alternative hypothesis, the excess Sr could lie initially on the nominally Ti-terminated \STO\ surface as clusters of residual atoms (possibly close to step edges) not removed by the surface treatment nominally guaranteeing the single TiO$_2$ termination. This would suggest the existence of a driving force tending to maintain, even in the presence of a non-uniform \STO\ termination,  a uniform SrO-TiO$_2$-LaO-AlO$_2$ sequence across the whole interface, by pushing the initial excess Sr to the top of the growing film. Finally, a finite amount of substrate surface Sr may be set free during the deposition process and experience an energy gain by floating at the surface of the heterostructure.

As for the BE shift between the two Sr components, it would be very tempting to attribute it to the electric potential foreseen to build-up across \LAO, within the polar catastrophe scenario. Nevertheless the analysis of current literature suggests that a BE shift of pure chemical nature, rather than of electrostatic nature, might well be at play. Chemical shifts very similar to the ones reported above have been in fact reported for bare, thermally treated \STO\ substrates and assigned to either SrO$_x$, \cite{Szot2000} or Sr bonded to carbon (e.g.\ SrCO$_3$). \cite{vanderHeide2001, Vasquez1991,Menou2009} A formation of Sr(OH)$_2$, due to the reaction with water, can also cause a similar BE shift. \cite{Vasquez1991}  This hypothesis, if confirmed, could be consistent with the observation that BE shifts of electrostatic nature are hardly found in \LAOS. \cite{Yoshimatsu2008,Segal2009,Takizawa2011,Slooten2013} 

At the end we address the general importance of the observed Sr-segregation for the physics of the oxide heterostructures. Any attempt to directly link this phenomenon to interface conductivity would be highly speculative at this stage. Migration of positively charged Sr$^{2+}$ atoms from the interface to the upper \LAO\ surface would certainly be an alternative or complementary means (with respect to electronic reconstruction) to alleviate the polar catastrophe. Nevertheless this argument would apply to any of the four cation species present in the system. Furthermore, this effect could be neutralized, if the migration involved neutral  Sr$^{2+}$ - O$^{2-}$ complexes rather than single ions. Finally, the very small amount of Sr-2 makes it insufficient to compensate a nominal polarity of 1/2 e$^-$ per in-plane uc. 
We will limit ourselves therefore to observe that the decade-long debate  on the origin of the 2DEG in \LAO/\STO\ has taught us that the finest details in the atomic arrangement, including e.g. the \STO\ atomic termination, or submonolayer differences in \LAO\ thickness above 3 uc, or \LAO\ stoichiometry variations of the order of 1$\%$ \cite{Warusawithana2013}, or undetectably low levels of oxygen vacancies, can dramatically alter the electronic properties of this system. Only a very accurate and complete understanding of the effective atomic configuration occurring in real systems will allow us to properly discern intrinsic and extrinsic effects. 

\section{Conclusions}

By varying the surface sensitivity of X-ray photoemission we have unambiguously identified a previously elusive high binding energy Sr-component that we attribute to a submonolayer thick overlayer. The BE shift can be assigned to a purely chemical shift between the covering layer, presumably SrO, SrCO$_3$ or Sr(OH)$_2$, and \STO, without any electrostatic contribution by the polar layer. The formation of such layer occurs both in \LAOS\ and in \NGOS\ for all thicknesses of the polar film. Our findings add further insight on the complex picture of oxide heterostructures, both in terms of their growth mechanisms and, possibly, of their electronic properties. They also confirm the tendency towards surface segregation 
of Sr in oxide systems with perovskite-related structure.

\section{Acknowledgement}
E.D.G., U.S.d.U., and F.M.G. acknowledge financial support by the European Union (Programme No.\ FP7/2007-2013, Grant Agreement No.\ 264098 – MAMA), and by the Ministero dell'Istruzione, dell'Universit\`{a} e della Ricerca (Grant No.\ PRIN 2010-11 – OXIDE).

%\bibliography{Paper2_rev2_final}

\begin{thebibliography}{39}
\bibitem [{\citenamefont {{Substrate related spurious Si intensity has been
  observed for low $\lambda$. The Si\,$2p$ line partly superimposes the
  La\,$4d$ line but can be removed by fitting with sufficient
  accuracy}}()}]{Note}%
  \BibitemOpen
  \bibfield  {author} {\bibinfo {author} {\bibnamefont {{Substrate related
  spurious Si intensity has been observed for low $\lambda$. The Si\,$2p$ line
  partly superimposes the La\,$4d$ line but can be removed by fitting with
  sufficient accuracy}}},\ }\href@noop {} {}\BibitemShut {NoStop}%
\end{thebibliography}


\end{document}